# Origin of gamma-ray families accompanied by halos and detected in experiments with x-ray emulsion chambers

Puchkov V.S. and Pyatovsky S.E.[1]

*Lebedev Physical Institute, Russian Academy of Sciences, Leninskii pr. 53, Moscow, 119991 Russia*

ABSTRACT – The phenomenon of gamma-ray families featuring halos that is observed in an experiment with x-ray emulsion chambers (XREC) in the Pamir experiment and in other XREC experiments is explained. The experimental properties of halos are analyzed via a comparison with the results of their simulation. It is shown that gamma-ray families featuring halos are predominantly produced (more than 96 % of them) by protons and heliumnuclei. This makes it possible to employ the experimental properties of halos to estimate the fraction of protons and helium nuclei in the mass composition of primary cosmic radiation.

1. INTRODUCTION

Since the early 1970s, the method of x-ray emulsion chambers (XREC) has widely been used to study cores of extensive air showers (EAS) at energies of primary particles in the range of $E_0$=1-100 PeV. New phenomena that could not be explained within the standard model of nuclear interactions were discovered experimentally on the basis of the XREC method. They include gamma-ray families featuring large-area halos, coplanar (aligned) events, penetrating hadrons characterized by an anomalously small absorption in lead, and centauro events [1]. Experiments based on the XREC method were performed at an altitude of 4400 m above sea level (a.s.l.) in the PAMIR mountains (Pamir XREC experiment) and at altitudes of 5200 m a.s.l. by the Brazil–Japan Collaboration at the Mount Chakaltaya, 5600 m a.s.l. by the China-Japan Collaboration at the Mount Kanbala, and 3800 m a.s.l. by the Japanese Interuniversity Collaboration at the Mount Fuji.

In the Pamir experiment, XREC is a solid-state tracking chamber that consists of absorber (lead and carbon) layers and x-ray-film layers for detecting the electron–photon cascade and the hadron component in an EAS core. The layout of the Pamir XREC setup is shown in Fig. 1. Within an air target above XREC (594 g/cm$^2$), nuclei of primary cosmic radiation initiate a nuclear-electromagnetic cascade whose highenergy ($E_\gamma \geq 1$ TeV) electron–photon component ($e^\pm$ and γ) is detected in the form of dark spots on an x-ray film, which are referred to as "γ rays" [2].

A high resolution of a two-sided x-ray film (about 100 $\mu$m) and a high photon-detection threshold (about 1 TeV) make it possible to detect, in XREC, narrow collimated beams of high-energy gamma rays in the form of groups of dark spots that are associated with individual EASs having identical zenith ($\theta$) and azimuthal ($\varphi$) angles and which are referred to as gamma-ray families. The accuracy in determining the coordinates and angles in the Pamir XREC experiment is $\Delta x, y \sim 100\ \mu$m, $\Delta \theta \sim 3°$, and $\Delta \varphi \sim 15°$. The photon-energy-weighted center of gamma-ray families corresponds to the coordinates of the EAS core to a precision not poorer than 1 cm. Thus, the XREC method permits visually detecting events occurring within a few centimeters from the EAS core.

---

[1] vgsep@ya.ru; sep@lebedev.ru

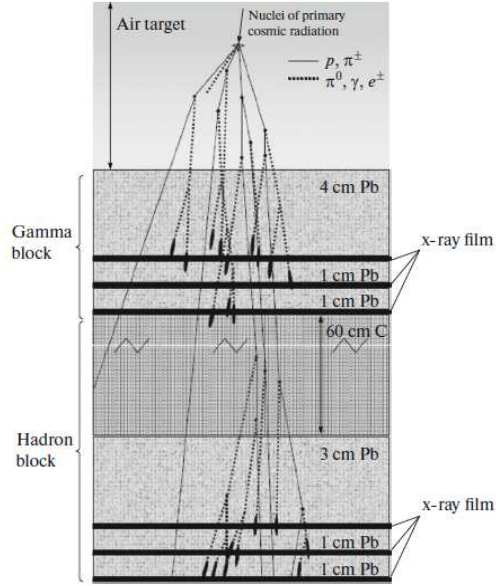

**Fig. 1.** Layout of the Pamir XREC (X-ray emulsion chamber) experiment.

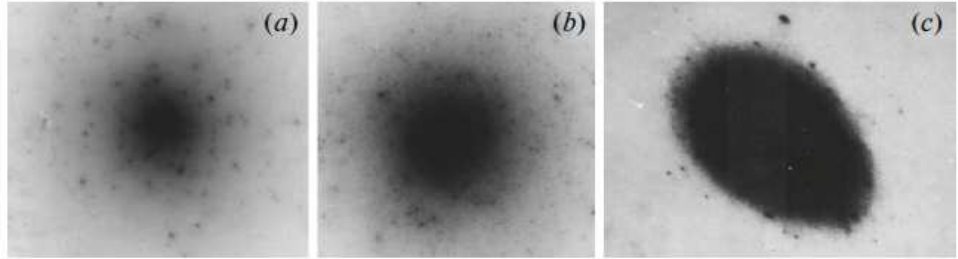

**Fig. 2.** Scanned image of the experimental halos (*a*) FIANIT ($S_{halo}$=1020 mm$^2$), (*b*) Tajikistan ($S_{halo}$=1451 mm$^2$), and (*c*) Andromeda ($S_{halo}$=950 mm$^2$).

The photon energy is determined from the photometry of dark spots on an x-ray film. The optical density of dark spots is proportional to the number of particles, $n=f(r,t,E)$ μm$^{-2}$, where $r$ is the distance that is measured from $e^{\pm}$ and $\gamma$ in EAS to the scanned cell and which is fitted between 1 μm and 6 cm, $t$ is the lead depth in the XREC gamma block with allowance for $\theta$ (see Fig. 1) (changing from 1 to 60 rad. units), and $E$ is the energy of $e^{\pm}$ and $\gamma$ in EASs (from 1 TeV to 1 EeV). The particle flux $n$ is proportional to the energy $E_\gamma$ of gamma rays that initiated a given cascade; that is,

$$D(n)=D_\infty(1-e^{-nS}), \tag{1}$$

where $S=(3.25\pm0.13)$ μm$^2$ is the effective emulsion grain area and $D_\infty=4$ is the maximum darkening of the x-ray film in the Pamir XREC experiment. In order to perform an absolute calibration of the XREC method for determining $E_\gamma$ from the darkening density $D$ on an x-ray film, use was made of the method for determining the $\pi^0$ mass by employing the measured energies $E_1$ and $E_2$ of the photons originating from the decay $\pi^0 \rightarrow 2\gamma$ and the laboratory opening angle between them, $\Phi$: $m_{\pi^0}=2\sin\Phi/2 \sqrt{E_1 E_2}$. This method makes it possible to assess the systematic error and to determine $E_\gamma$ to a precision in the range of $\sigma(E_\gamma)/E_\gamma \sim 0.2$-$0.3$.

As $\Sigma E_\gamma$ for the gamma-ray families grows ($E_\gamma \geq 4$ TeV), regions of high optical density that are referred to as halos and which are characterized by an area $S$ of a few units of to a few thousand millimeters squared appear on an x-ray film in XREC at the center of gamma-ray families. The halo fraction in the total amount of gamma-ray families grows with $\Sigma E_\gamma$ and, for $\Sigma E_\gamma \geq 1$ PeV, becomes as large as about 70 % of all gamma-ray families.

For the first time, the halo phenomenon was detected by the Brazil–Japan Collaboration at the Mount Chacaltaya Laboratory. The event in question was classed with exotic nuclear-interaction events not described within the standard model of nuclear interactions and was called Andromeda (Fig. 2). Similar events were observed in the Pamir XREC experiment (Fianit and Tajikistan halos – see Fig. 2).

Respective calculations permitted obtaining halos from single gamma rays, but the areas of these halos did not exceed 40 mm². The area spectrum of calculated halos did not agree with the area spectrum of experimental halos. In order to describe the halo phenomenon and to obtain a halo of area in excess of 100 mm², it was necessary to take into account lateral distribution functions specially calculated for the Pamir experiment and generated in lead by beams of gamma rays produced via nuclear interactions in the atmosphere above the chamber, as well as to include additionally subthreshold gamma rays of energy in the range of $E_\gamma \geq 100$ GeV. With allowance for the overlap of the lateral distribution functions and subthreshold gamma rays, halos of area equal to or larger than 1000 mm² were obtained on the basis of the standard model of nuclear interactions without resort to a change in the primary interaction event or to exotic particles.

The air target above XREC has a large thickness, so that several generations of particles from the nuclear-electromagnetic cascade contribute to gamma-ray families observed in XREC. This requires taking into account the XREC response in analyzing experimental data. The experimental data from the Pamir XREC experiment were analyzed on the basis of the MC0 model formulated in [3], which was specially developed for the Pamir XREC experiment. The MC0 model provides good agreement of the results of calculations with data from the Pamir XREC experiment and with accelerator data at lower energies. This model was developed with the aim of simulating hadron-hadron and hadron-nucleus interactions – in particular, high-$x_F$ experimental data near the EAS core at distances of up to about 50 cm. It permits describing all properties of gamma-ray families obtained in the Pamir XPEC experiment. This could not be done within models involving a slower energy dissipation in the course of the development of nuclear-electromagnetic cascades.

**Table 1.** Fraction of multicenter γ-ray families with halos generated by various nuclei of primary cosmic radiation

| $P$ | He | C | Fe | MC0 | PAMIR |
|---|---|---|---|---|---|
| 0.25±0.03 | 0.45±0.09 | 0.59±0.11 | 0.70±0.03 | 0.28±0.03 | 0.23±0.07 |

At the present time, there are no models that would describe simultaneously XREC and EAS experimental data. A slower energy dissipation in the development of nuclear-electromagnetic cascades in the atmosphere (QGSJet 01/II-03/II-04, SIBYLL 2.1/2.3c, EPOS 1.99/LHC) entails an increase in the average radius of γ-ray families and in their structurality, which is given in Table 1; according to the simulation, it is 0.28±0.03, the value obtained for it experimentally being 0.23±0.07. From the results presented in Table 1, it follows that one cannot enhance the fraction of structured halos within the model, since it would not comply with the experimental value. However, even calculations based on models that describe EASs lead to the conclusion that only events initiated by primary protons, possibly involving a fraction of helium nuclei, are detected in the Pamir XREC experiment.

With the advent new accelerator data, MC0 has been being corrected [3]. The averaged inelasticinteraction cross section obtained in various experiments at the LHC with a large spread and large errors and used in calculations is $\sigma_{inel}$(7 TeV)=68.0±2.0(syst.)±2.4(lumi.)±4.0(extrap.) mb. Thus, the cross section in the MC0 model and models that describe EAS agree within the errors and do not contradict experimental data. Agreement between the model data and the LHC results on multiplicities is within 10 %. It is also noteworthy that XREC experiments detects gamma-ray families in the region of $x_F \geq 0.05$. Data in this region are scanty and come only from the LHCf (LHC forward) experiment at pseudorapidities in narrow ranges. The parameters of gamma-ray families in XREC are determined by transverse momenta and $x_F$ values in the range where accelerator data are scarce.

2. SIMULATION OF HALOS

According to the results of a simulation, gammaray families featuring a halo are observed starting from $E_0 \sim 5$ PeV (Fig. 3). The simulation reveals that, at this energy, the lateral distributions of gamma rays in lead begin to overlap one another, forming regions of high optical density (Fig. 2). According to investigations, a halo can also be formed by a single photon initiated by a neutral pion. In that case, however, the halo area would not exceed about 40 mm² for a standard atmosphere, but this is substantially smaller than the halo areas observed in the Pamir XREC experiment and in other XREC experiments. The bounded height of the atmosphere prevents us from obtaining calculated halos from a single photon with areas that would agree with the experimental areas.

The calculations performed in the present study permitted reproducing the experimentally observed area spectrum of halos (see Fig. 4), obtaining gammaray families that have a halo and which possess properties complying with their experimental counterparts, and explaining the origin of halos within the standard model of nuclear interactions.

The simulation of gamma-ray families at the Pamir XREC observation level in the primary-cosmicradiation energy range of $E_0$=0.1-3000 PeV was performed for the following criteria corresponding to the Pamir experiment:

(i) The energy threshold for EAS particles is 100 GeV.

(ii) The minimum total energy of EAS particles at the observation level is 100 TeV.

(iii) Primary cosmic radiation is formed by groups of $p$, He, C, N, O, Mg, Si, V, and Fe nuclei.

(iv) The observation-level depth in the Pamir XREC experiment is 594 g/cm$^2$.

(v) The arrival angle $\theta$ of primary-cosmic-radiation nuclei changes isotropically between 0 and 0.9 rad.

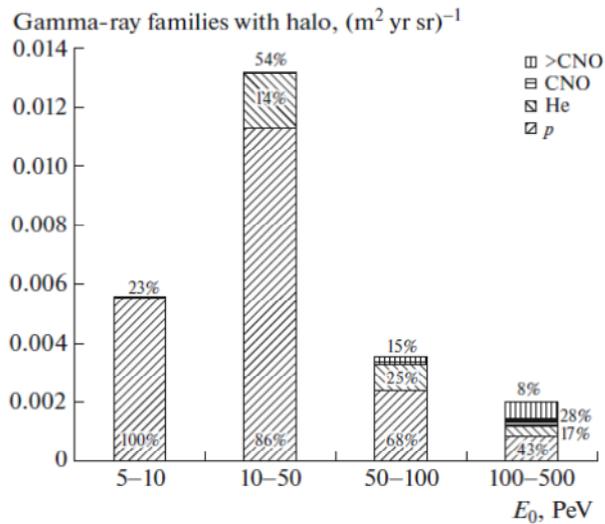

**Fig. 3.** $E_0$ dependence of the intensity of $\gamma$-ray families with halos and relative fractions of primary-cosmicradiation nuclei producing these halos according to the simulation.

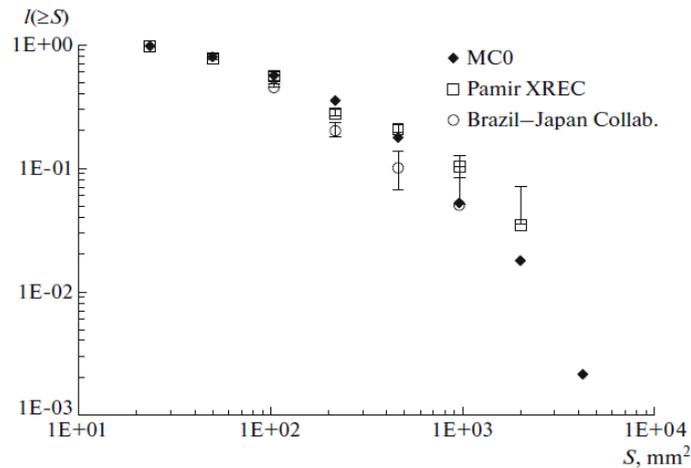

**Fig. 4.** Area spectrum of the halos obtained in the Pamir XREC experiment and the experiments of the Brazil–Japan Collaboration along with the halo area spectrum from the simulation.

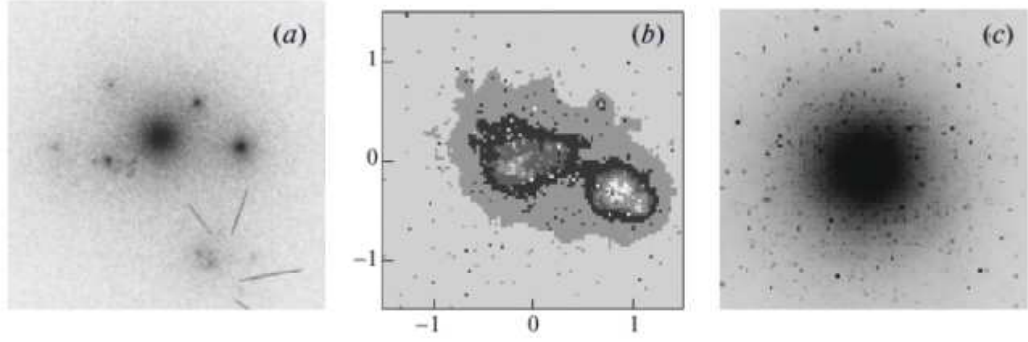

**Fig. 5.** (*a*) Scanned image of a multicenter gamma-ray family featuring a halo from the Pamir XREC experiment ($\Sigma E_\gamma$=1688 TeV and $S_{halo}$=46 mm$^2$); (*b*) calculated gamma-ray family characterized by the presence of a halo and produced by a primary helium nucleus ($\Sigma E_\gamma$=1296 TeV and $S_{halo}$=67 mm$^2$), a *D* isogram being given; and (*c*) calculated halo formed by a primary proton ($\Sigma E_\gamma$=39.5 PeV and $S_{halo}$=2100 mm$^2$).

The gamma-ray families featuring a halo were simulated by employing EASs generated by the Monte Carlo method, and the halo selection criteria in the Pamir XREC experiments (Table 2) [4] were taken into account in this simulation.

Specifically, these criteria are the following:

(i) The halo area $S_{D=0.5}$ is bounded by the $D$=0.5 optical-density isoline, which corresponds to the particle-flux density of 0.04 $\mu$m$^{-2}$: $S_{D=0.5} \geq 4$ mm$^2$ for a one-center halo and $\Sigma S_{i\ D=0.5} \geq 4$ mm$^2$ in the case of $S_{i\ D=0.5} \geq 1$ mm$^2$ for multicenter halos. Examples of experimental and calculated multicenter gamma-ray families featuring a halo are given in Fig. 5 along with a calculated large-area halo.

(ii) The depth in XREC of the x-ray film in which gamma-ray families featuring a halo are detected is 5 cm of lead, which is 9 to 11 rad. units, depending on the EAS arrival angle $\theta$.

(iii) The number of gamma-ray families with $\Sigma E_\gamma \geq 400$ TeV in the Pamir XREC experiment is 143; 61 of them have a halo [5].

In the chessboard algorithm used in simulating gamma-ray families with halos, use is made of the photon lateral distributions in lead that were specially calculated for the Pamir XREC experiment. These lateral distributions make it possible imitate the propagation of gamma rays through XREC, provide a good description of the development of the electron–photon cascade in XREC at large distances with allowance for the multilayered structure of the Pamir XREC array, and take into account the Landau-Pomeranchuk-Migdal effect.

**Table 2.** Criteria of simulation and selection of EASs and gamma-ray families with halos

| Simulation step | Observation level in Pamir XREC gamma block | Particles at observation level | Minimum total particle energy, TeV | Distance from particles to EAS axis, cm | Lead depth in XREC | Halo area, mm$^2$ |
|---|---|---|---|---|---|---|
| 1. Calculation of EAS according to MC0 | Above lead absorber | All EAS | 100 | All EAS particles | - | - |
| 2. Selection of EAS according to Pamir XREC criteria | Above lead absorber | $e^\pm, \gamma$ | 100 | All EAS particles | - | - |

| Simulation step | Observation level in Pamir XREC gamma block | Particles at observation level | Minimum total particle energy, TeV | Distance from particles to EAS axis, cm | Lead depth in XREC | Halo area, mm² |
|---|---|---|---|---|---|---|
| 3. Selection of gamma-ray families for calculations according to Pamir XREC criteria | Above lead absorber | $e^{\pm}, \gamma$ | $\Sigma E_\gamma \geq 400$ TeV ($E_\gamma \geq 4$ TeV) | $\leq 15$ | - | - |
| 4. Selection of γ-ray families c $S$ according to Pamir XREC criteria | X-ray film | $e^{\pm}, \gamma$ | $\Sigma E_\gamma \geq 400$ TeV ($E_\gamma \geq 4$ TeV) | $\leq 15$ | 5 cm/$\cos\theta$ | $\Sigma S_i \geq 4$, where $S_i \geq 1$, $D \geq 0.5$ |

The chessboard algorithm imitates the operation of a scanner in the photometry of experimentally observed gamma-ray families. The image of gammaray families with halos on an x-ray film is scanned with a step of 100 μm, and the darkening level $D$ of all 100×100 μm cells of the halo being studied is analyzed. In simulating gamma-ray families featuring a halo, the known particle flux $n$ obtained from the lateral distribution determining the dependence $n=f(r, t, E)$ is transformed into $D$ according to Eq. (1).

Figures 6 and 7 show the calculated gamma-ray families characterized by the presence of a halo and generated by primary protons, helium nuclei and iron nuclei. One can see peaks corresponding to individual photons and having $D \geq 0.5$ but $S_i < 1$ mm².

3. PROPERTIES OF HALOS

ThePamir XREC experiment collected 61 gamma-ray families featuring halos with $\Sigma S_i \geq 4$ mm² upon the exposure of 3000 m² yr sr, while the experiments at Chakaltaya, Kanbala, and Fuji observed 20 gamma-ray families featuring halos with $\Sigma S_i \geq 10$ mm². More than 600 gamma-ray families with halos from the simulation were analyzed.

Basic properties of the gamma-ray families with halos in the Pamir XREC experiment include the halo area $S_{halo}$; the structurality of halos (number of centers with $S_i \geq 1$ mm²) (see Fig. 7; see also Table 1); and the energy of gamma-ray families, $\Sigma E_\gamma$ ($E_\gamma \geq 4$ TeV), as well as the calculated threshold in $E_0$ for halo formation (Fig. 3) and the probability for halo formation by various nuclei of primary cosmic radiation (Table 3).

3.1. Halo areas

The area spectrum of the halos detected in the Pamir XREC experiment and in the experiments of the Brazil-Japan Collaboration is shown in Fig. 4. Good agreement between the calculated and experimental spectra ($R^2 > 95$ %) indicates that, within the standard model of nuclear interactions, we can obtain halos of both small and large area and that halos originate from the overlap of the lateral distributions of $e^{\pm}$ and γ. The simulation reveals that the contribution of subthreshold gamma rays with energy below 100 GeV to $S_{halo}$ is about 3 to 5 % for halos of area $S_{halo} \sim (4\text{-}10)$ mm² and about 5 to 10 % for halos of area $S_{halo} \sim 100$ mm². This was taken into account in the present data analysis.

### 3.2. Halo structure

In a number cases, gamma-ray families featuring a halo consist of several centers [6] (see Figs. 5 and 7). The fraction of multicenter halos is a parameter that depends strongly on the sort of primary-cosmicradiation nuclei (Table 1).

From the data in Table 1, it follows that the standard model of nuclear interactions describes well the halo-structurality parameter. Halos are predominantly produced by primary protons, possibly with a small admixture of helium nuclei. This conclusion is confirmed by the probabilities for the formation of gamma-ray families with halos by primary-cosmicradiation nuclei of energy in the region of $E_0 \geq 5$ PeV (Table 3).

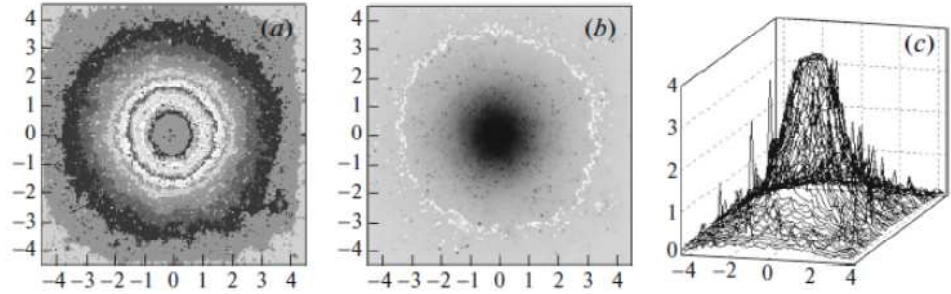

**Fig. 6.** Calculated one-center halo that was produced by a primary iron nucleus ($E_0$=1.63 EeV and $S_{halo}$=3783 mm$^2$): (*a*) *D* isogram; (*b*) model image on an x-ray film, *D*=0.5 boundary being indicated by a white contour; and (*c*) three-dimensional isogram.

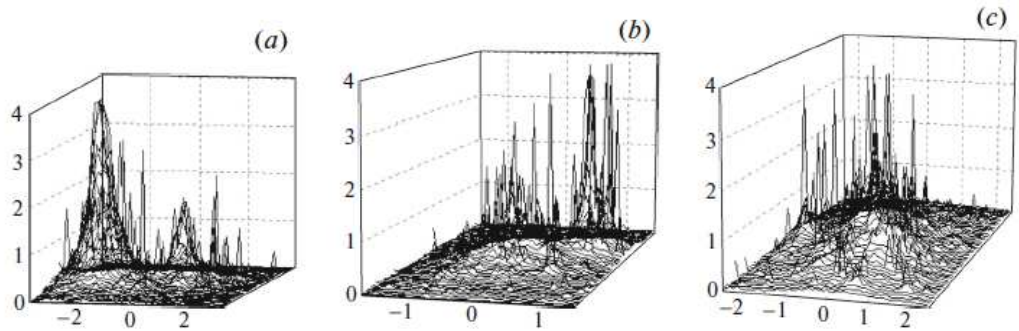

**Fig. 7.** Calculated multicenter gamma-ray families characterized by the presence of a halo and produced by (*a*) a primary proton ($E_0$=0.19 EeV, $S_1$=529 mm$^2$, and $S_2$=102 mm$^2$), (*b*) a primary helium nucleus ($E_0$=0.07 EeV, $S_1$=40 mm$^2$, and $S_2$=27 mm$^2$), and (*c*) a primary iron nucleus ($E_0$=0.69 EeV, $S_1$=441 mm$^2$, and $S_2$=21 mm$^2$).

### 3.3. Probabilities for halo formation

The calculations on the basis of the MC0 model [3] involved a normalization to the experimental total spectrum of primary cosmic radiation with respect to $E_0$. According to these calculations, more than 83 %, about 13 %, and about 4 % of all gamma-ray families featuring halos are produced by, respectively, protons, helium nuclei, and heavy nuclei (see Fig. 3). As is shown in Table 3, the probability for the formation of gamma-ray families with halos by protons is four times as great as the probability for their formation by helium nuclei. The contribution of heavy nuclei is about 4 %, which makes it possible to estimate the fraction of protons and helium nuclei in the mass composition of primary cosmic radiation. Tables 1 and 3 show the formation of experimentally observed gamma-ray families featuring a halo is due predominantly to protons.

**Table 3.** Probabilities (in percent) for the formation of gamma-ray familieswith halos by nuclei of primary cosmic radiation

| *p* | He | >He |
|---|---|---|
| 1.76±0.01 | 0.44±0.02 | 0.13±0.02 |

At energies as high as $E_0 \sim 1$ EeV, the probability for the formation of gamma-ray families with halos by all sorts of primary-cosmic-radiation nuclei is close to 100 %, but the contribution of primary-cosmicradiation nuclei with energies around $E_0 \sim 1$ EeV to the total number of events is insignificant in view of the descending power-law character of the energy spectrum of primary cosmic radiation. Thus, the gamma-ray families featuring a halo are predominantly produced by protons and helium nuclei, as can be seen from Fig. 3.

## 4. ESTIMATION OF THE FRACTIONS OF PROTONS AND HELIUM NUCLEI IN THE MASS COMPOSITION OF PRIMARY COSMIC RADIATION

The fractions of protons and helium nuclei in the mass composition of primary cosmic radiation is a hotly debated issue. For example, two aspects of it were discussed in [7-9] and were also addressed in the Tunka [10, 11] and Tunka-Rex [12] experiments and in the studies of other collaborations. These were an individual estimation of the fraction of protons and the fraction of helium nuclei in the mass composition of primary-cosmic radiation and the question of whether the mass composition of primary cosmic radiation becomes heavier or lighter in the knee region. Figure 8 shown the change in the fraction of protons and the fraction of helium nuclei in the range of $E_0=1$-$100$ PeV according to data from the Pamir, KASCADE [8, 13-15], ARGO-YBJ [16], Tunka [17-21], and IceCube [22] experiments. The results of the Pamir XREC experiment that concern the estimation of the fraction of protons and the fraction of helium nuclei in the mass composition of primary cosmic radiation at $E_0 \sim 10$ PeV on the basis of an analysis of the gamma-ray families featuring a halo showed good agreement with the results of the Tunka and IceCube experiments. The reason behind the discrepancy between the Pamir XREC results, on one hand, and the KASCADE and ARGO-YBJ data, on the other hand, calls for a dedicated consideration.

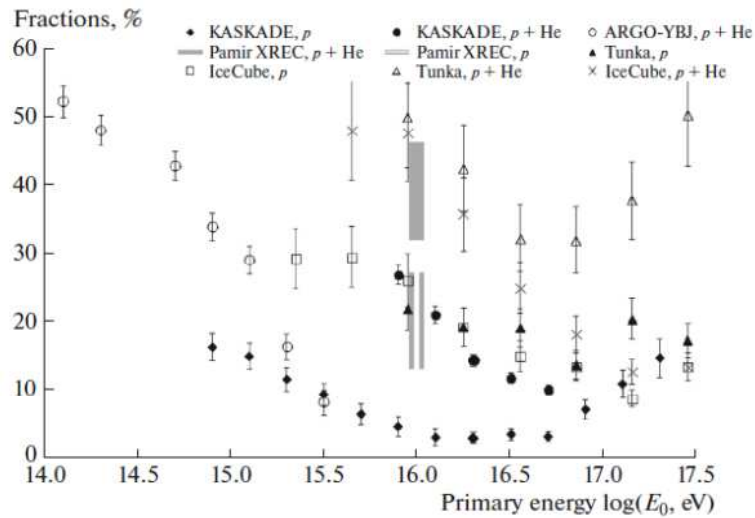

**Fig. 8.** Proton and $p$+He fractions in the mass composition of primary cosmic radiation according to data of the KASCADE [8, 13-15], ARGO-YBJ [16], Tunka [17-21], IceCube [22], and Pamir experiments.

The probabilities for the formation of gamma-ray families with halos by protons and helium nuclei differ by a factor of four. On the basis of the known numbers of experimentally observed gamma-ray families featuring halos, one can deduce the fraction of protons and the fraction of helium nuclei in the mass composition of primary cosmic radiation at the weightedmean energy of nuclei that produced halo detected in the Pamir XREC experiment, about 10 PeV.

The results of the simulation revealed that, if primary cosmic radiation consisted of protons exclusively, then the number of observed gamma-ray families featuring a halo would have been 140 instead of 61±8 ones obtained experimentally. For helium and heavier nuclei, the number of observed gammaray families featuring halos should be 34 and 5, respectively. Approximately equal fractions of protons and helium nuclei in the mass composition of primary cosmic radiation correspond to the experimental number of gamma-ray families with halos and to the known probabilities for the formation of gamma-ray families with halos in Table 3. These fractions are (20±6) % at $E_0 \sim 10$ PeV. The value obtained in the Pamir XREC experiment for the fraction of protons and helium nuclei in the mass composition of primary cosmic radiation is greater than the $p$+He fraction estimated in the KASCADE experiment, about 25 % at $E_0 \sim 10$ PeV, and in the ARGO-YBJ experiment, not greater than 10 % at $E_0=3$ PeV.

## 5. CONCLUSIONS

Quantitative estimates obtained from analyses of EAS features are model-dependent, which refers to the majority of experiments. However, the conclusions drawn in the present study from the results of the Pamir XREC experiment are virtually modelindependent. For example, the origin of large-area halos, which was explained by the overlap of the lateral distributions of photon beams and by the inclusion of subthreshold photons of energy not lower than 100 GeV, will be obtained by means of any model, since it is associated with the propagation of the electron–photon cascade through the XREC gamma block.

Also, the EAS features showing the highest sensitivity to the mass composition are detected and analyzed near the EAS axis at distances of several centimeters from it. This accuracy of EAS axis localization is possible within the XREC method but is inaccessible in EAS experiments. The most recent publications on EASs demonstrate that the model accuracy of EAS axis localization is up to 1 m. At thesame time, one can visually observe, on an x-ray film of XREC, gamma-ray families produced by protons and helium nuclei and thereby localize the EAS axis to a precision of about 1 cm. At the present time, it is impossible to observe visually this region either in the LHCf experiment or in experiments with resistive counters.

The foregoing also applies to the probability for the formation of halos by nuclei of different sorts. It is of importance that the probability for the formation of halos by protons is four times as high as the probability of their formation by helium nuclei. This makes it possible to estimate individually the fractions of protons and helium nuclei in primary cosmic radiation. This ratio of the probabilities would stem from the application of anymodel; otherwise, themodels would not describe EAS properties.

The calculations performed in the present study have explained the origin of gamma-ray families featuring a halo within the standard model of nuclear interactions, which provides a good description of experimental data from the Pamir XREC array:

(i) The halo phenomenon in gamma-ray families is not an exotic event but is an overlap of individual dark spots from the electron–photon cascade on an x-ray film.

(ii) The gamma-ray families detected in XREC that are sensitive to the mass composition of primary cosmic radiation lie near the EAS axis at a distance of up to 50 cm, the accuracy of EAS axis localization in the Pamir XREC experiment being about 1 cm. In XREC experiments, the EAS axis is visually observed as an image on an x-ray film in the form of gammaray families.

(iii) The probability for the formation of gamma-ray families with halos by protons is four times as great as the probability of their formation by helium nuclei. Other nuclei make virtually no contribution to the formation of gamma-ray families, and this enables one to estimate the fraction of protons and the fraction of helium nuclei in the mass composition of primary cosmic radiation. Either fraction as obtained on the basis of the experimental number of halos, 61±8, is (20±6) % at $E_0 \sim 10$ PeV. This value is greater than the analogous estimates based on data from the KASCADE and ARGO-YBJ EAS experiments and comply with the estimates in the experiments of the Tunka and IceCube Collaborations.